\documentclass[twocolumn,aps,prb,superscriptaddress,amsmath,amssymb,superscriptaddress]{revtex4-1}
\newcommand{\pagenumbaa}{1}
\usepackage{graphicx}
\usepackage{color}
\usepackage{braket}

\usepackage{verbatim}
\usepackage{amsmath}
\usepackage{mathtools}

\usepackage{float}
\usepackage{hyperref}
\usepackage{lipsum}

\begin{document}
\title{Optically driving the radiative Auger transition}

\author{Clemens Spinnler\textsuperscript{1,*}}
\author{Liang Zhai\textsuperscript{1,*}}
\author{Giang N. Nguyen\textsuperscript{1}}
\author{Julian Ritzmann\textsuperscript{2}}
\author{Andreas D. Wieck\textsuperscript{2}}
\author{Arne Ludwig\textsuperscript{2}}
\author{Alisa Javadi\textsuperscript{1}}
\author{Doris E. Reiter\textsuperscript{4}}
\author{Pawe\l{} Machnikowski\textsuperscript{3}}
\author{Richard J. Warburton\textsuperscript{1}}
\author{Matthias C. L\"obl\textsuperscript{1,*,$\dagger$}}

\noaffiliation
\affiliation{Department of Physics, University of Basel, Klingelbergstrasse 82, 4056 Basel, Switzerland}
\affiliation{Lehrstuhl f\"ur Angewandte Festk\"orperphysik, Ruhr-Universit\"at Bochum, 44780 Bochum, Germany}
\affiliation{Department of Theoretical Physics, Wroc\l{}aw University of Science and Technology, 50-370 Wroc\l{}aw, Poland}
\affiliation{Institut f\"ur Festk\"orpertheorie, Universit\"at M\"unster, 48149 M\"unster, Germany
\\ \ \\
\textsuperscript{*}\,These authors contributed equally to this work.
\\
\textsuperscript{$\dagger$}\,Correspondence should be addressed to: matthias.loebl@unibas.ch
\\ \ \\
}

\begin{abstract}
In a radiative Auger process, optical decay is accompanied by simultaneous excitation of other carriers \cite{Bloch1935}. The radiative Auger process gives rise to weak red-shifted satellite peaks in the optical emission spectrum \cite{Hawrylak1991b,Llusar2020}. These satellite peaks have been observed over a large spectral range: in the X-ray emission of atoms \cite{Aberg1969}; close to visible frequencies on donors in semiconductors \cite{Dean1967} and quantum emitters \cite{Lobl2020,Antolinez2019}; and at infrared frequencies as shake-up lines in two-dimensional systems \cite{Nash1993,Skolnick1994,Finkelstein1997,Manfra1998}. So far, all the work on the radiative Auger process has focussed on detecting the spontaneous emission. However, the fact that the radiative Auger process leads to photon emission suggests that the transition can also be optically excited. In such an inverted radiative Auger process, excitation would correspond to simultaneous photon absorption and electronic de-excitation. Here, we demonstrate optical driving of the radiative Auger transition on a trion in a semiconductor quantum dot. The radiative Auger and the fundamental transition together form a $\Lambda$-system \cite{Fleischhauer2005}. On driving both transitions of this $\Lambda$-system simultaneously, we observe a reduction of the fluorescence signal by up to $70\%$. Our results demonstrate a type of optically addressable transition connecting few-body Coulomb interactions to quantum optics. The results open up the possibility of carrying out THz spectroscopy on single quantum emitters with all the benefits of optics: coherent laser sources, efficient and fast single-photon detectors. In analogy to optical control of an electron spin \cite{Press2008}, the $\Lambda$-system between the radiative Auger and the fundamental transitions allows optical control of the emitters' orbital degree of freedom.
\end{abstract}

\maketitle

\setcounter{page}{\pagenumbaa}
\thispagestyle{plain}
The radiative Auger process (shake-up) arises from Coulomb interactions between charged particles in close proximity \cite{Bloch1935,Llusar2020}. When one electron loses energy by spontaneous photon emission, these interactions change in a sudden manner. As a consequence, other electrons can be promoted into higher shells and the emitted photon is red-shifted \cite{Stohr2013}. Further, radiative Auger connects carrier dynamics to the quantum optical properties of the emitted photons \cite{Lobl2020}, making it a powerful probe of multi-particle systems. While driving the fundamental transition between the electron ground state and an optically excited state is an important technique in quantum optics \cite{Muller2007,Flagg2009}, driving the radiative Auger transition has not been achieved. Success here would significantly increase the number of quantum optics techniques that can be employed.

\begin{figure*}[t]
\includegraphics[width=1.8\columnwidth]{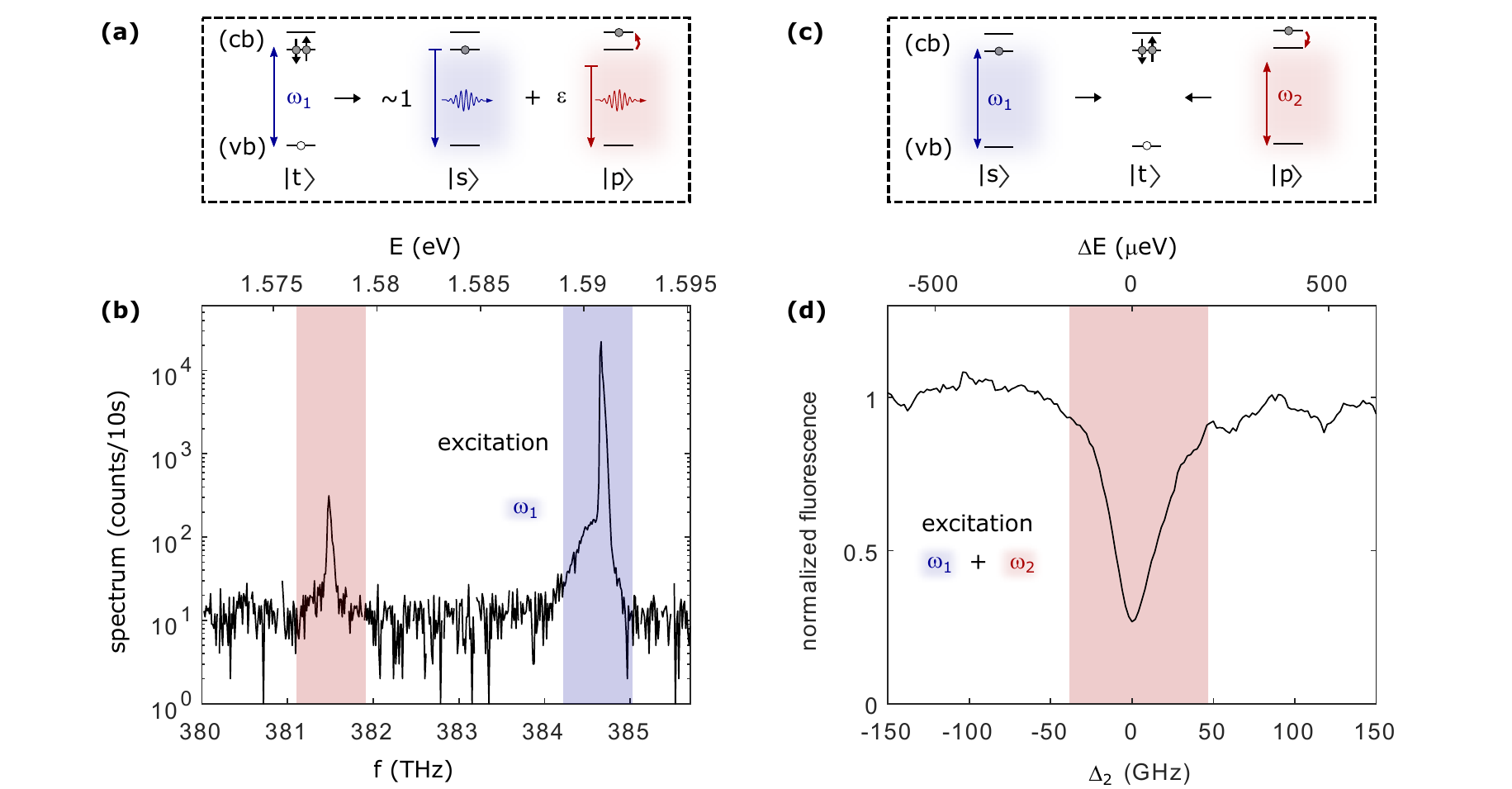}
\caption{\label{fig:intro}\textbf{Radiative Auger emission and excitation of the radiative Auger transition. (a)} Schematic illustration of the fundamental transition and the radiative Auger process. The trion state $\ket{t}$ optically decays by recombination of one electron from conduction band (cb) with a hole from the valence band (vb). The second electron either stays in its ground state $\ket{s}$ (fundamental transition), or is promoted to a higher shell $\ket{p}$ (radiative Auger). The radiative Auger photon is red-shifted from the fundamental transition by the energy transferred to the Auger electron. \textbf{(b)} Emission spectrum from a negatively charged quantum dot upon optical excitation at the fundamental transition. In addition to the fundamental transition (highlighted in blue), there is a red-shifted satellite line (highlighted in red). This emission arises from the radiative Auger process where the trion state $\ket{t}$ decays to the excited electron state $\ket{p}$. \textbf{(c)} Two possible absorption channels in the presence of one confined conduction band electron. When the electron is in the ground state $\ket{s}$, a laser resonant with the fundamental transition (blue, frequency $\omega_1$) excites a valence band electron and brings the system to the trion state, $\ket{t}$. When the conduction band electron is in an excited state $\ket{p}$, a red-shifted laser (frequency $\omega_2$) can excite the system to the same trion state $\ket{t}$. In this inverted radiative Auger process, the missing energy is provided by the excited electron. \textbf{(d)} Resonance fluorescence from the fundamental transition in the presence of a strong second laser. When the second laser ($\omega_2$) is on resonance with the radiative Auger transition ($\Delta_2=0$), the resonance fluorescence intensity is strongly reduced.}
\end{figure*}

We demonstrate driving the radiative Auger transition on a GaAs quantum dot embedded in AlGaAs \cite{Wang2007,Zhai2019}. The quantum dot forms a potential minimum and confines charge carriers, resulting in discrete energy levels like in an atom. Without optical illumination, a single electron is trapped in the conduction band of the quantum dot and occupies the lowest possible shell (the $s$-shell, $\ket{s}$). Upon resonant excitation of the fundamental transition, a second electron is promoted from the filled valence band to the conduction band and a negative trion $X^{1-}$ ($\ket{t}$) is formed. This trion consists of two electrons in the conduction band and one electron-vacancy (hole) in the valence band. Fig.\ \ref{fig:intro}(a) shows the possible optical decay paths: in the fundamental transition, one electron decays removing the valence band hole while the other electron remains in the conduction band ground state $\ket{s}$; in the radiative Auger process, the remaining electron is promoted into an excited state $\ket{p}$. The emitted photon is red-shifted by the energy separation between $\ket{p}$ and $\ket{s}$ \cite{Bloch1935,Lobl2020}. Fig.\ \ref{fig:intro}(b) shows a typical emission spectrum from the trion decay. This spectrum is measured on resonantly driving the fundamental transition $\ket{s}$--$\ket{t}$ (at $384.7\ \text{THz}\ \hat{=}\ 1.591\ \text{eV}$) with a narrow-bandwidth laser \cite{Lobl2020}. Red-shifted by $3.2\ \text{THz}$ ($13.2\ \text{meV}$) from the fundamental transition there is a weak satellite line that arises from the radiative Auger process.

Photons at the radiative Auger frequency have insufficient energy to excite the fundamental transition $\ket{s}$--$\ket{t}$. Figure\ \ref{fig:intro}(c) shows how the trion state $\ket{t}$ still can be excited with a laser at the Auger transition. The missing energy is provided by the electron which initially occupies the excited state $\ket{p}$. However, driving the radiative Auger transition is experimentally challenging mainly for two reasons: first, there is a fast non-radiative relaxation from the excited single-electron state $\ket{p}$ back to $\ket{s}$ \cite{Zibik2009,Lobl2020}, and the state $\ket{p}$ is not occupied at thermal equilibrium. Second, the dipole moment of the radiative Auger transition is small. Therefore, it is difficult to achieve high Rabi frequencies on driving the transition, plus the radiative Auger emission is very weak and hard to distinguish from the back-reflected laser light.

We solve these issues by driving the fundamental transition $\ket{s}$--$\ket{t}$ with one laser (labelled as $\omega_1$) while simultaneously driving the radiative Auger transition with a second laser (labelled as $\omega_2$). This $\Lambda$-configuration has the following advantages: first, on driving $\ket{s}$--$\ket{t}$ with $\omega_1$, there is a small chance of initializing the system in state $\ket{p}$ via the radiative Auger emission. Additionally, driving the $\ket{p}$--$\ket{t}$-transition with $\omega_2$, while transferring population to $\ket{t}$ with $\omega_1$, also leads to a finite occupation of $\ket{p}$. Second, the small dipole matrix element of the radiative Auger transition is compensated by using high power for $\omega_2$. The high power causes a high laser background when detecting the fluorescence from the radiative Auger transition. Instead, we tune the second laser over the Auger transition while measuring just the fluorescence originating from the fundamental transition $\ket{s}$--$\ket{t}$. Fig.\ \ref{fig:intro}(d) shows the result of this two-laser experiment. We observe a strong reduction in fluorescence on addressing the transition $\ket{p}$--$\ket{t}$ which is characteristic for two-colour excitation of a $\Lambda$-configuration. Our approach has a conceptual similarity to the driving of weak phonon sidebands of mechanical resonators resulting in optomechanically induced transparency \cite{Weis2010,Safavi2011}.

\begin{figure*}[t]
\includegraphics[width=1.8\columnwidth]{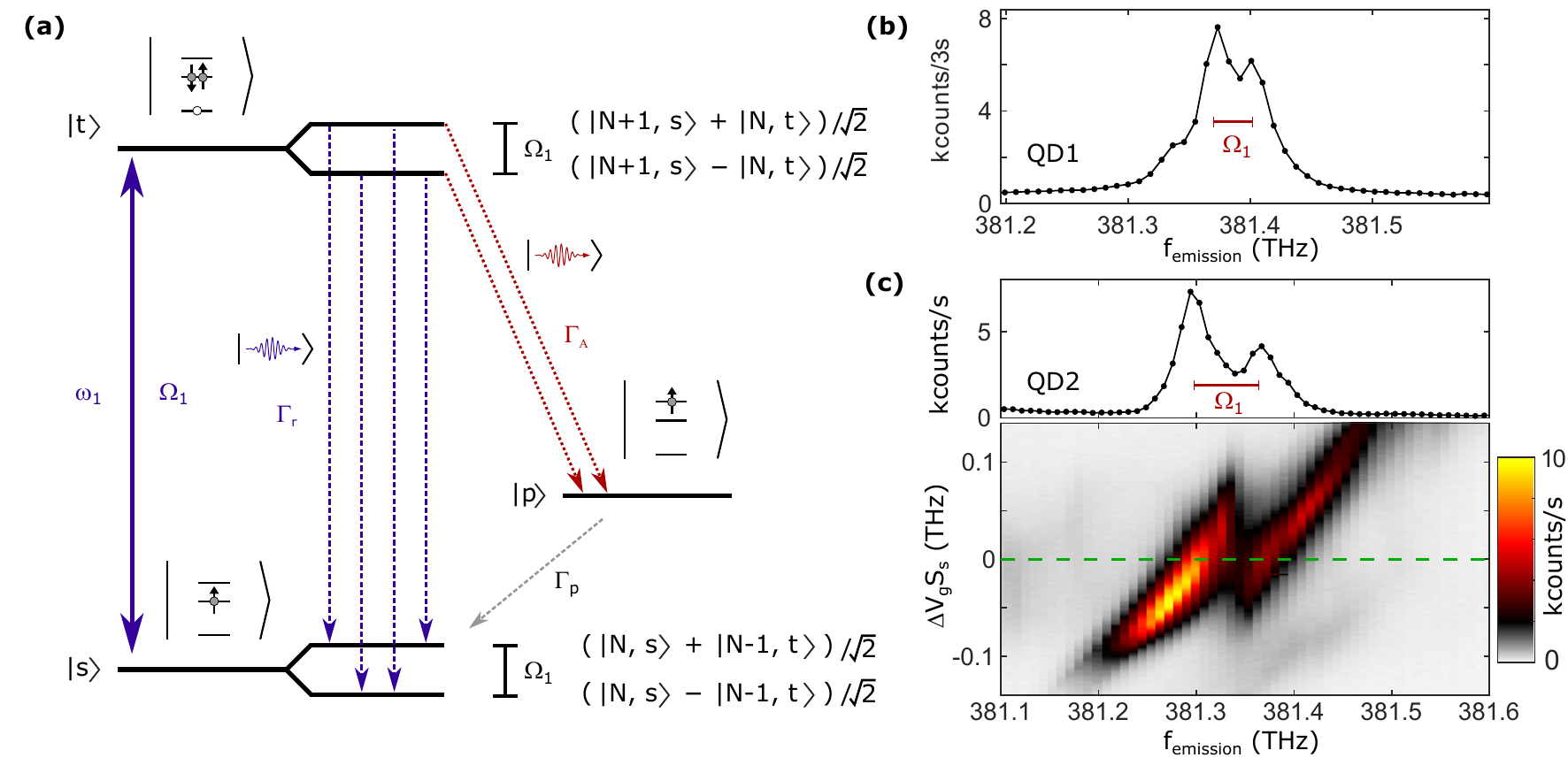}
\caption{\label{fig:AutlerTownes}\textbf{Autler-Townes splitting in the radiative Auger emission. (a)} Level scheme under strong resonant driving of the fundamental transition ($\ket{s}$--$\ket{t}$). The energy levels of the transition split into dressed states. The splitting between the dressed states is given by the Rabi frequency, $\Omega_1$. In the radiative Auger emission (red arrows), the dressed-state splitting enables two decay paths leading to an Autler-Townes splitting. \textbf{(b)} Radiative Auger emission from the main quantum dot (QD1) on driving the transition $\ket{s}$--$\ket{t}$ with $\omega_1$. The Rabi frequency of $\Omega_1=2\pi\times31.9\ \text{GHz}$ (red bar) is determined from a power saturation curve. The measured Autler-Townes splitting in the emission matches the Rabi frequency. \textbf{(c)} Emission spectrum from a second quantum dot (QD2), measured for a set of different detunings ($\Delta_1=\Delta V_g\cdot S_s$) between fundamental transition and laser. The upper part of the plot is a line cut along the dashed green line at zero detuning ($\Delta_1=0$). In this case, the measured Autler-Townes splitting agrees with the independently determined Rabi frequency ($\Omega_1=2\pi\times67.7\ $GHz).}
\end{figure*}

\begin{figure*}
\includegraphics[width=1.8\columnwidth]{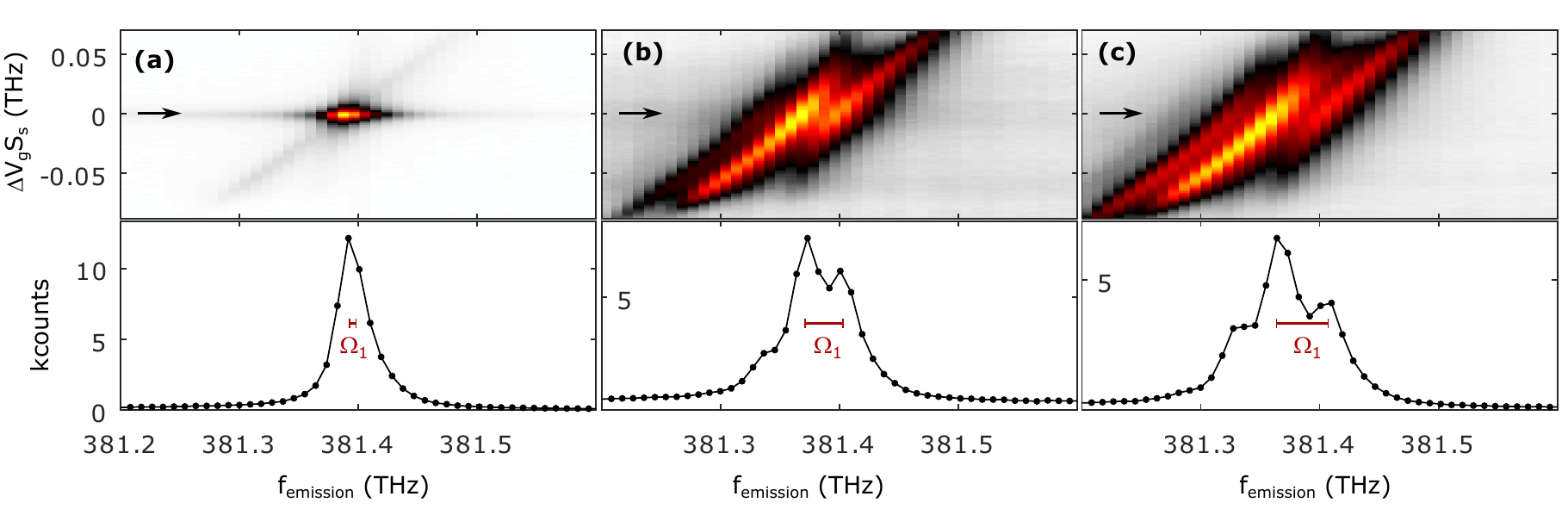}
\caption{\label{fig:AutlerTownes_suppl}\textbf{Radiative Auger emission upon excitation of the fundamental transition. (a)} Radiative Auger emission from QD1 (see also Fig.\ \ref{fig:AutlerTownes}(b)). The emission frequency (x-axis) is plotted as a function of detuning between laser and the $\ket{s}-\ket{t}$ transition (y-axis). The quantum dot transitions are detuned from the fixed laser by applying a gate voltage, $V_g$. The detuning from the fundamental transition is $\Delta V\cdot S_s$, where $S_s$ is the Stark-shift of the fundamental transition and $\Delta V_g$ the difference in gate voltage. In the measurements shown here, the Autler-Townes splitting is not resolved since the Rabi frequency is too small ($\Omega_1=2\pi\times5.5\ \text{GHz}$). On detuning the quantum dot resonance from the laser ($\Delta V_g\neq0$), there is a small probability to excite the trion via the phonon sideband giving rise to a weak ``diagonal" emission line. In the case of a red-detuned quantum dot ($\Delta V_g<0$), the laser has more energy than the quantum dot transition and the additional energy can be transferred to LA-phonons. In the case of a blue-detuned quantum dot, the laser energy is too little and the missing energy can be provided by phonon absorption. \textbf{(b, c)} The same measurements as before performed at higher Rabi frequencies ($\Omega_1=2\pi\times31.9\ \text{GHz}$, $\Omega_1=2\pi\times43.2\ \text{GHz}$) where the Autler-Townes splittings are resolved. The Autler-Townes splittings are independently determined from a power saturation curve (red bars). They match the measured splittings in the emission spectra.}
\end{figure*}

We consider initially the situation where the fundamental transition ($\ket{s}$--$\ket{t}$) is strongly driven by a single laser. If radiative Auger and fundamental transition form a $\Lambda$-system, one would expect an Autler-Townes splitting in the radiative Auger emission. Fig.\ \ref{fig:AutlerTownes}(a) shows the corresponding level scheme including the dressed states $\frac{1}{\sqrt{2}}(\ket{N+1,s}\pm\ket{N,t})$ and $\frac{1}{\sqrt{2}}(\ket{N,s}\pm\ket{N-1,t})$, where $N$ is the photon number. The dressed-state splitting leads to the Mollow triplet in the resonance fluorescence \cite{Mollow1969,Flagg2009,Ulhaq2012}. For a decay into a third level, the Autler-Townes splitting \cite{Autler1955,Xu2007} in the emission is expected to be the Rabi frequency $\Omega_1$. Fig.\ \ref{fig:AutlerTownes}(b) shows the radiative Auger emission of one quantum dot (QD1). In this measurement, the laser is on resonance with the fundamental transition. The Rabi frequency ($\Omega_1=2\pi\times31.9\ $GHz, red bar in Fig.\ \ref{fig:AutlerTownes}(b)) is estimated independently by measuring the fluorescence intensity as a function of laser power (Methods Fig.\ \ref{fig:Bfield}(b)). We observe an Autler-Townes splitting that agrees well with this Rabi frequency. For this quantum dot, we also observe an additional weak emission appearing on the low energy side of the spectrum when using high Rabi frequencies (see Fig.\ \ref{fig:AutlerTownes}(b) and Methods Fig.\ \ref{fig:AutlerTownes_suppl}). We speculate that this emission is connected to optical coupling between $\ket{p}$ and an excited trion state, $\ket{t^*}$. Figure\ \ref{fig:AutlerTownes}(c) shows radiative Auger emission from a second quantum dot (QD2). For this quantum dot, we measure the radiative Auger emission as a function of detuning between the quantum dot transition and the laser (see Methods Fig.\ \ref{fig:AutlerTownes_suppl} for the corresponding measurement on QD1). On applying a gate voltage $\Delta V_g$, the quantum dot transition $\ket{s}$--$\ket{t}$ is detuned from the fixed laser by $\Delta_1=\Delta V_g\cdot S_s$ via the quantum-confined Stark shift. $S_s$ parameterizes the Stark shift of the fundamental transition. At zero detuning, the observed Autler-Townes splitting again agrees with the Rabi frequency obtained from a power saturation curve ($\Omega_1=2\pi\times67.7\ $GHz).

\begin{figure*}[t]
\includegraphics[width=1.8\columnwidth]{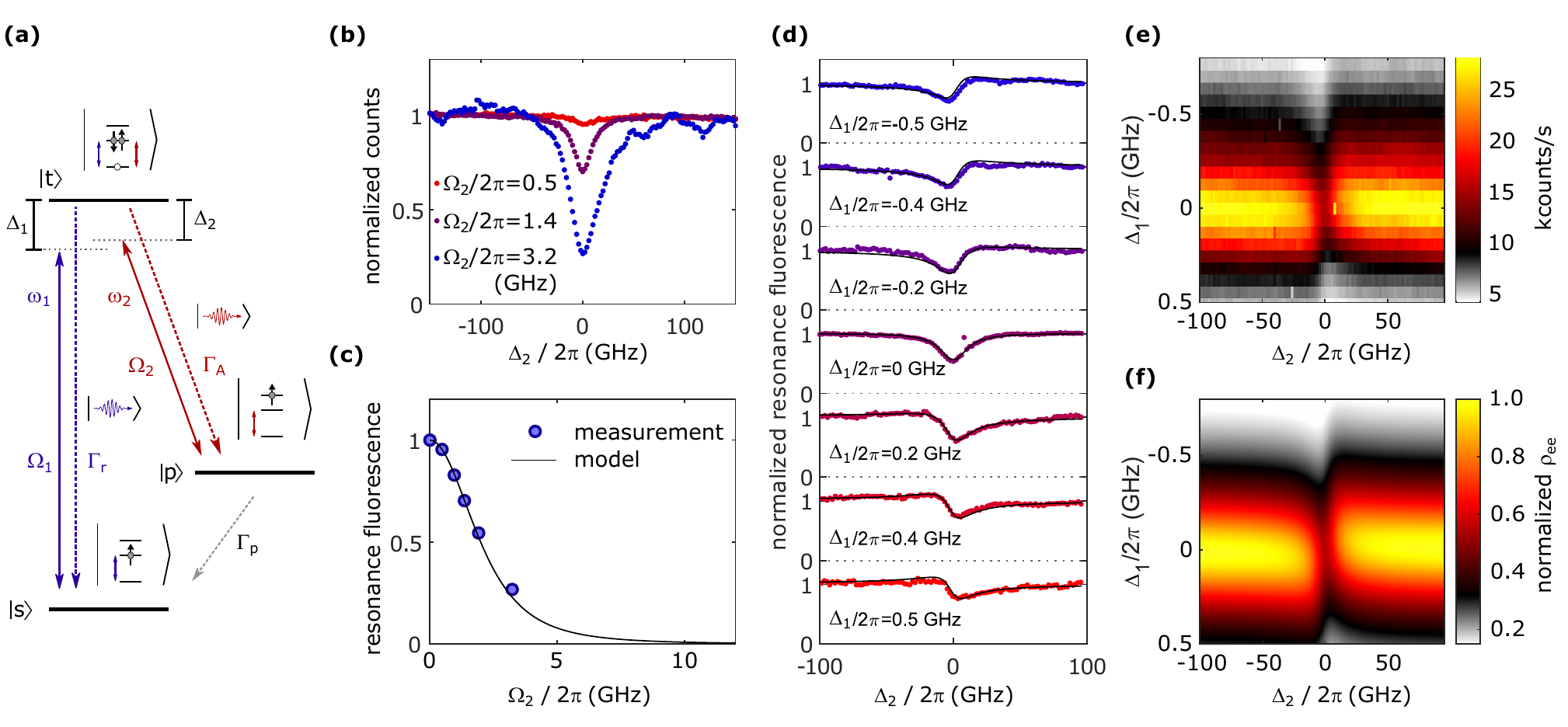}
\caption{\label{fig:driveAuger}\textbf{Optically driving the radiative Auger transition. (a)} The level scheme where one laser ($\omega_1$) with Rabi frequency $\Omega_1$ drives the fundamental transition ($\ket{s}$--$\ket{t}$) while a second laser ($\omega_2$) drives the radiative Auger transition ($\ket{p}$--$\ket{t}$) with Rabi frequency $\Omega_2$. \textbf{(b)} Resonance fluorescence as a function of detuning $\Delta_2$ (detuning of $\omega_2$). At low values of $\Omega_2$, the resonance fluorescence intensity is almost constant for different values of $\Delta_2$. For the highest value of $\Omega_2$, the resonance fluorescence drops by up to $\sim70\%$ on bringing $\omega_2$ into resonance with the radiative Auger transition. The strong fluorescence dip at a particular frequency is a characteristic feature of a $\Lambda$-system driven with two lasers that are detuned in frequency by the ground state splitting. \textbf{(c)} Resonance fluorescence at $\Delta_2=0$ as a function of $\Omega_2$. The resonance fluorescence intensity (blue dots) drops with increasing $\Omega_2$, fitting well to the theoretical model (black line). \textbf{(d)} Fluorescence intensity as a function of detuning $\Delta_2$. The Rabi frequencies are $\Omega_1=2\pi\times0.27\ \text{GHz}$, $\Omega_2=2\pi\times2.1\ \text{GHz}$. The same measurement is repeated for a series of fixed detunings $\Delta_1$ (detuning of $\omega_1$ from the fundamental transition). Detuning $\omega_1$ leads to an asymmetric fluorescence dip. This asymmetry is well captured by our quantum optics simulations (black lines) based on the level scheme shown in (a). \textbf{(e)} Fluorescence intensity as a function of laser detunings $\Delta_1$, $\Delta_2$. \textbf{(f)} Simulation of the fluorescence intensity as a function of the laser detunings.}
\end{figure*}

We now consider the experiments with the second laser (labelled as $\omega_2$) at the radiative Auger transition. Fig.\ \ref{fig:driveAuger}(a) shows the corresponding level scheme. We set $\omega_1$ to a modest Rabi frequency ($\Omega_1=2\pi\times0.08\ \text{GHz}$) compared to the decay rate of the trion ($\Gamma_r=2\pi\times0.50\ \text{GHz}$). The frequency of the radiative Auger transition is estimated from the trion emission spectrum (Fig.\ \ref{fig:intro}(b)). We sweep the frequency $\omega_2$ and simultaneously monitor the resonance fluorescence intensity from the fundamental transition. Fig.\ \ref{fig:driveAuger}(b) shows this measurement for different Rabi frequencies $\Omega_2$. On increasing the power of $\omega_2$ to several orders of magnitude higher than the power of $\omega_1$, there is a pronounced dip in the fluorescence intensity. This intensity dip appears precisely when the laser frequency $\omega_2$ matches the radiative Auger transition ($\ket{p}$--$\ket{t}$) and is characteristic for a $\Lambda$-system that is driven with two lasers. We estimate the Rabi frequency $\Omega_2$ driving $\ket{p}$--$\ket{t}$ by simulating the resonance fluorescence intensity as a function of $\Delta_2$ (see Supplement for the quantum optics simulation). In this simulation we keep the decay rate from $\ket{p}$ to $\ket{s}$ ($\Gamma_p\sim2\pi\times9.3\ \text{GHz}$) fixed to the value that we determine from independent auto- and cross-correlation measurements \cite{Lobl2020} (see Methods Fig.\ \ref{fig:g2}(d)). Additionally, we fit a constant pure dephasing, $\gamma_p$, for the state $\ket{p}$ which leads to an additional broadening of the fluorescence dip. We estimate $\gamma_p\sim2\pi\times8.8\ \text{GHz}$ from the fit and a Rabi frequency of $\Omega_2=2\pi\times3.2\ \text{GHz}$ ($\omega_2$) for the strongest fluorescence dip. Note that additional excitation-induced dephasing via phonons is expected to be weak for such Rabi-frequencies \cite{Ramsay2010,Reiter2019}.

In Fig.\ \ref{fig:driveAuger}(c), we plot the minimum of the resonance fluorescence dip as a function of $\Omega_2$. The data fits well to the $\Lambda$-system model with two driving lasers. For the highest value of $\Omega_2$, we achieve a reduction of the resonance fluorescence intensity by up to $70\%$. The intensity reduction is limited by the power that we can reach in our optical setup. The measurement shows that resonance fluorescence can be switched on and off by using the radiative Auger transition. In our system, part of the fluorescence dip is due to reduction of the overall absorption via the formation of a dark state. This effect is related to electromagnetically induced transparency (EIT) \cite{Fleischhauer2005} and coherent population trapping (CPT) \cite{Aspect1988,Prechtel2016}. An additional reduction of the signal comes from the fact that there is a fast decay rate $\Gamma_p$ from state $\ket{p}$ to $\ket{s}$. Thus, after the laser-induced transition from state $\ket{t}$ to $\ket{p}$, the system quickly decays to the ground state $\ket{s}$. This de-excitation channel reduces the population of the trion state and therefore the fluorescence intensity.

The measurements so far were performed with $\omega_1$ on resonance ($\Delta_1=0$). We repeat the two-laser experiments while detuning $\omega_1$ from the fundamental transition. Fig.\ \ref{fig:driveAuger}(d) shows the fluorescence intensity for positive, zero, and negative detuning $\Delta_1$. For non-zero detuning, the fluorescence dip is asymmetric as a function of $\Delta_2$. The asymmetry is an important result as it cannot be explained by a rate equation description, but depends on quantum coherences in a master equation model (see Supplement). The full dependence of the resonance fluorescence intensity as a function of $\Delta_1$ and $\Delta_2$ is plotted in Fig.\ \ref{fig:driveAuger}(e). This data set fits well to the corresponding quantum optics simulation in Fig.\ \ref{fig:driveAuger}(f) using the parameters from the previous measurements.\newline

In summary, we have shown that it is possible to drive the radiative Auger transition. The resonance fluorescence can be strongly reduced by addressing this transition: a modulated laser on the radiative Auger transition could be used for fast optical gating of the emitter's absorption. As an outlook, we suggest that an effective coupling between the orbital states, split by frequencies in the THz band, can be created by two lasers at optical frequencies. The idea here is to establish a Raman-like process: the lasers are equally detuned from their resonances and an exciton is not created. This scheme facilitates control of the orbital degree of freedom with techniques that have been developed for manipulating spin-states \cite{Press2008,Prechtel2016}. Further quantum optics experiments with radiative Auger photons are conceivable: by using a two-colour Raman-scheme \cite{Gangloff2019}, it might be possible to create deterministically highly excited shake-up states that are also subject of recent theoretical investigations \cite{Bauman2020}. Adding a third laser with a THz-frequency at the transition $\ket{s}$--$\ket{p}$ \cite{Zibik2009} might even allow close-contour driving schemes \cite{Barfuss2018}. In analogy to experiments on spins \cite{Gao2012}, the radiative Auger process could lead to an entanglement between the frequency of the emitted photon and the orbital state of the Auger electron. 

\section*{Author Contributions}
\label{sec:contrib}
JR, LZ, ADW, AL grew the sample. CS, LZ, GNN fabricated the sample. LZ, MCL, JR, AL designed the sample. LZ, CS, GNN, MCL carried out the experiments. MCL, CS, PM, DER, LZ, GNN, AJ, RJW analyzed the data and developed the theory. CS and MCL prepared the various figures. MCL initiated and conceived the project. MCL and RJW wrote the manuscript with input from all the authors.

\section*{Acknowledgments}
\label{sec:acknowledgement}
We thank Krzysztof Gawarecki and Philipp Treutlein for fruitful discussions. We acknowledge financial support from Swiss National Science Foundation project 200020\_175748, NCCR QSIT and Horizon-2020 FET-Open Project QLUSTER. LZ received funding from the Marie Sk\l{}odowska-Curie grant agreement No.\ 721394 (4PHOTON). GNN received funding from the Marie Sk\l{}odowska-Curie Grant Agreement No.\ 861097 (QUDOT-TECH). AJ received funding from the Marie Sk\l{}odowska-Curie grant agreement No.\ 840453 (HiFig). JR, AL, and ADW gratefully acknowledge financial support from the grants DFH/UFA CDFA05-06, DFG TRR160, DFG project 383065199, and BMBF Q.Link.X 16KIS0867.

\section*{Methods}
\label{sec:methods}
\renewcommand{\figurename}{Methods FIG.}

\begin{figure*}
\includegraphics[width=1.8\columnwidth]{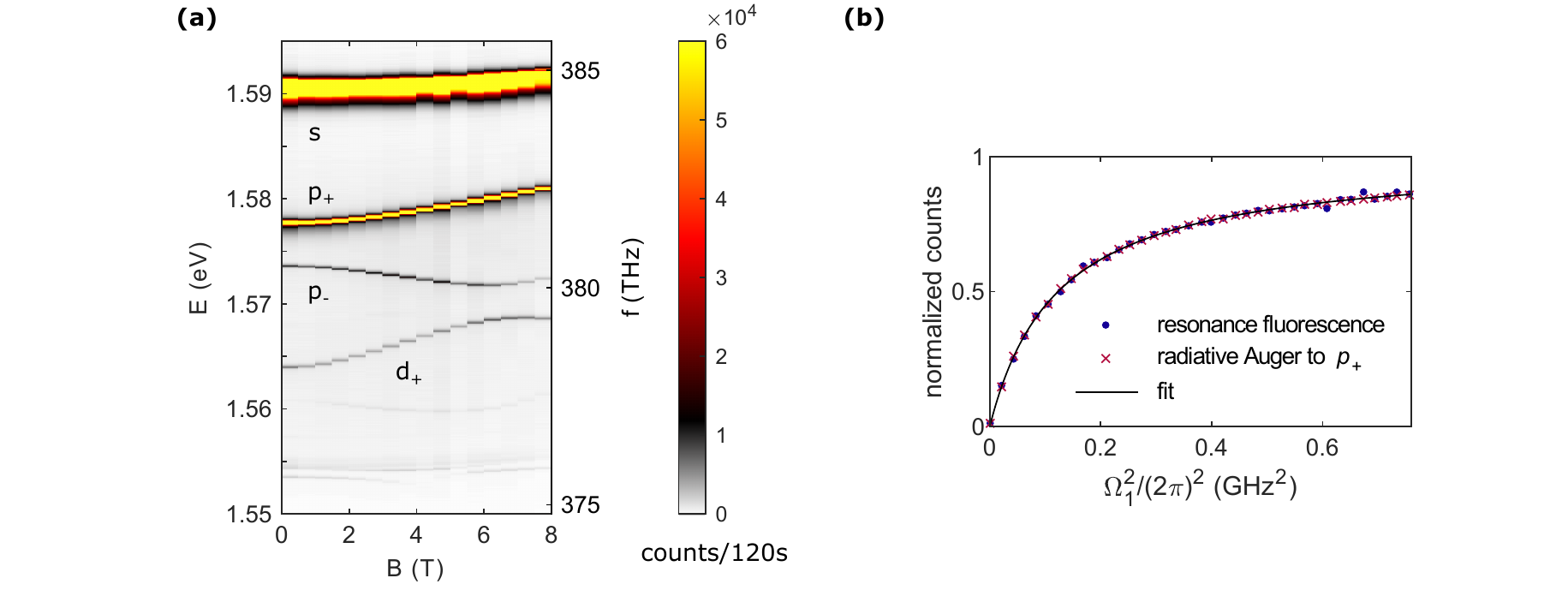}
\caption{\label{fig:Bfield}\textbf{Magnetic field dependence of the different emission lines. (a)} Resonance fluorescence from the fundamental transition and radiative Auger emission as a function of the magnetic field, $B$ (data from QD1). The strong magnetic field dispersion of the radiative Auger lines enables direct identification of the final electron states of the Auger electron. \textbf{(b)} Normalized fluorescence intensity of the different emission lines as a function of Rabi frequency $\Omega_1$. $\omega_2$ is turned off for this measurement ($\Omega_2=0$). The power dependence of the radiative Auger intensity coincides with that of the resonance fluorescence from the fundamental transition and matches the power curve of a two-level system. From a fit to the power curve we estimate $\Omega_1$ in our measurements. For fitting, the radiative decay is fixed to a value that we determine from an independent lifetime measurement ($\Gamma_r=2\pi\times0.50\ \text{GHz}$) \cite{Zhai2019}.}
\end{figure*}

For all our measurements, the quantum dot sample is kept in a liquid helium bath cryostat at 4.2 K. The quantum dots used in this work are GaAs quantum dots in AlGaAs. Their decay rates $\Gamma_r$ were determined by lifetime measurements using pulsed resonant excitation \cite{Zhai2019}. QD1 is identical to the second quantum dot in Ref.\ \onlinecite{Zhai2019}. The quantum dots presented in this work have a pronounced radiative Auger emission compared to other III-V quantum dots \cite{Lobl2020} indicating a stronger dipole-moment of the radiative Auger transition. We use radiative Auger lines where the final state of the Auger electron, $\ket{p}$, is a quantum dot $p$-shell. In particular, we investigate the transition associated with the lower $p$-shell ($p_+$) for QD1 and the higher $p$-shell ($p_-$) for QD2. We can assign further emission lines to the corresponding higher electronic shells by measuring the magnetic field dispersion of the emission spectrum \cite{Lobl2020} (see Methods Fig.\ \ref{fig:Bfield}(a)).

To excite the quantum dots, we use a tunable diode laser with a narrow bandwidth far below the quantum dot linewidth. Resonant excitation is not necessary to see the radiative Auger emission, above-band excitation is sufficient as well \citep{Antolinez2019,Llusar2020}. However, resonant excitation has the advantage that no continuum states are excited making it easier to identify all emission lines. For this work, resonant excitation is crucial to address optically a single radiative Auger transition. To suppress the reflected excitation laser, we use a cross-polarization scheme \cite{Kuhlmann2013a}.

\begin{figure*}
\includegraphics[width=1.8\columnwidth]{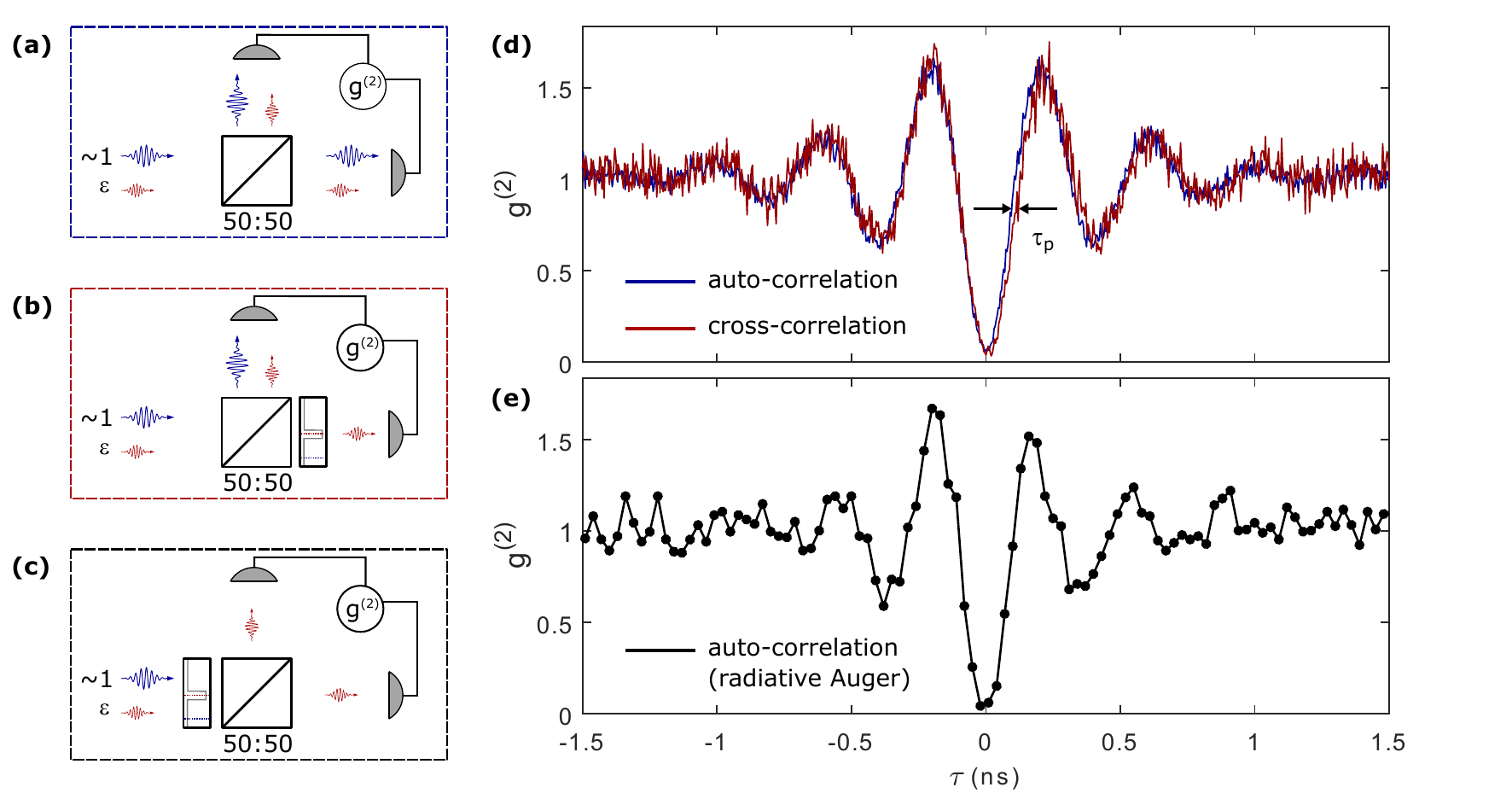}
\caption{\label{fig:g2}\textbf{Time-resolved correlation measurements. (a)} Schematic measurement setup for the auto-correlation of resonance fluorescence from the fundamental transition. \textbf{(b)} Schematic setup for the cross-correlation between resonance fluorescence from the fundamental transition and radiative Auger emission. \textbf{(c)} Schematic setup for the auto-correlation of the radiative Auger emission. \textbf{(d)} Comparison between the fundamental transition auto-correlation (blue) and the cross-correlation between resonance fluorescence from the fundamental transition and radiative Auger emission (data from QD1). Both $g^{(2)}$-measurements are performed with a single laser on the fundamental transition and show Rabi oscillations due to the strong driving ($\Omega_1$). The cross-correlation has a small offset, $\tau_p=1/\Gamma_p$, towards positive time delays. This offset measures the finite time for which the Auger electron remains in the excited state after a radiative Auger process has occurred \citep{Lobl2020}. The origin of the relaxation $\Gamma_p\sim2\pi\times9.3\ \text{GHz}$ is probably a phonon-assisted decay \cite{Zibik2009} but further investigations are needed. \textbf{(e)} Auto-correlation of the radiative Auger emission. Since the radiative Auger emission is relatively weak (count rates: $630\ $Hz on the first, $530\ $Hz on the second detector), a long integration time ($\sim50\ $h) is needed to resolve the Rabi oscillations in this measurement. The used Rabi frequency is slightly different with respect to the auto- and cross-correlation shown in (d).}
\end{figure*}

To determine the relaxation rate $\Gamma_p$ from $\ket{p}$ to $\ket{s}$, we make use of a technique developed in Ref.\ \onlinecite{Lobl2020}: on driving $\ket{s}$--$\ket{t}$ ($\Omega_2=0$), we measure an auto-correlation of the resonance fluorescence from the fundamental transition and compare it to the cross-correlation between resonance fluorescence from the fundamental transition and radiative Auger emission. The corresponding measurement setups are shown in Methods Fig.\ \ref{fig:g2}(a,b). To resolve the auto- and cross-correlations with high time resolution, we use two superconducting nanowire single-photon detectors (SingleQuantum) with a timing jitter below $20$ ps (FWHM) in combination with correlation hardware (Swabian Instruments).

Compared to the auto-correlation, the cross-correlation has a small time offset when a radiative Auger photon is followed by a photon from the fundamental transition (positive time delay in Methods Fig.\ \ref{fig:g2}(d)). This time scale corresponds to the relaxation time, $\tau_p=1/\Gamma_p$, describing the relaxation from $\ket{p}$ to $\ket{s}$. The relaxation time appears in the cross-correlation: when a radiative Auger event is detected by the first detector, there is an additional waiting time of $\tau_p$ before the excited Auger electron relaxes to the ground state and the system can be optically re-excited. Therefore, it takes longer before a second photon is detected. The additional waiting time is only present for the cross-correlation. For the auto-correlation, the system decays directly to the ground state $\ket{s}$ and there is no additional waiting time.

Finally, we also measure the auto-correlation of the radiative Auger emission (see Methods Fig.\ \ref{fig:g2}(c) for the setup). The measurement is shown in Methods Fig.\ \ref{fig:g2}(e). We observe a pronounced anti-bunching at zero delay proving the single-photon nature of the radiative Auger photons. Going beyond the results in Ref.\ \onlinecite{Lobl2020}, we observe the Rabi oscillation from strongly driving the transition $\ket{s}$--$\ket{t}$ in the photon-statistics of the radiative Auger photons from the transition $\ket{p}$--$\ket{t}$.

\bibliography{radAuger2_arxiv_version.bbl}

\end{document}